\documentclass{Interspeech2024}
\usepackage{multirow}  % 添加 multirow 宏包
\usepackage{afterpage}
\usepackage{float}
\interspeechcameraready

\title{Emotion-Aware Speech Self-Supervised Representation Learning with Intensity Knowledge}

\name[affiliation={1}]{Rui}{Liu}
\name[affiliation={1}]{Zening}{Ma}

\address{ $^1$ 
Inner Mongolia University, Hohhot, China  
\email{liurui\_imu@163.com, codening\_2022@163.com}
}


\begin{document}

\maketitle

% the abstract here must exactly match the abstract entered into the paper submission system
\begin{abstract}
    
% 1000 characters. ASCII characters only. No citations.
Speech Self-Supervised Learning (SSL) has demonstrated considerable efficacy in various downstream tasks. Nevertheless, prevailing self-supervised models often overlook the incorporation of emotion-related prior information, thereby neglecting the potential enhancement of emotion task comprehension through emotion prior knowledge in speech. In this paper, we propose an emotion-aware speech representation learning with intensity knowledge. Specifically, we extract frame-level emotion intensities using an established speech-emotion understanding model. Subsequently, we propose a novel emotional masking strategy (EMS) to incorporate emotion intensities into the masking process. We selected two representative models based on Transformer and CNN, namely MockingJay and Non-autoregressive Predictive Coding (NPC), and conducted experiments on IEMOCAP dataset. Experiments have demonstrated that the representations derived from our proposed method outperform the original model in SER task.
\end{abstract}

\section{Introduction}

Self-supervised learning (SSL) based pre-training models such as BERT~\cite{devlin2018bert}, GPT~\cite{radford2018improving}, ALBERT~\cite{lan2019albert}, and data2vec~\cite{baevski2022data2vec} have been instrumental in acquiring contextual information from large-scale unlabeled data through meticulously designed pre-training tasks. SSL models were initially used in Natural Language Processing (NLP) and have now been extended to the field of speech. The representations obtained using these models as feature extractors have proven to be very effective and yield excellent results when applied to speech and language processing (SLP) downstream tasks. Raw speech signals contain a wealth of acoustic and linguistic information and are adept at conveying speaker characteristics, emotions, and even intentions. However, extracting these high-level attributes from surface features such as log Mel spectrograms, Mel frequency cepstrum coefficients, or waveforms is a major challenge. Self-supervised modeling provides a solution that extracts high-level representations from these surface features, allowing downstream tasks to easily access the potential knowledge embedded in the original speech signal. 

Speech Emotion Recognition (SER) \cite{FAN2024102522,liu2024contrastive} is a crucial component of human-machine interaction. With the advancements in deep learning, some studies~\cite{fayek2017evaluating,vaswani2017attention} aim to learn emotion representations from audio signals using neural networks \cite{zuo2023exploiting}. However, compared to other common downstream tasks like Automatic Speech Recognition (ASR), SER datasets~\cite{busso2008iemocap,livingstone2018ryerson} are relatively smaller. Therefore, leveraging self-supervised pre-training models to learn representations from a large volume of unlabeled speech data and subsequently using these models either as feature extractors or by directly fine-tuning the entire model has become a common solution.

In the past few years, numerous researchers have delved into the study of self-supervised models to obtain advanced representations containing richer information. For instance, Mockingjay~\cite{liu2020mockingjay} proposed unsupervised training to learn speech representations without relying on any labels. This was achieved by employing multi-layer transformer encoders and multi-head self-attention~\cite{vaswani2017attention} to enable bidirectional encoding. The proposed Masked Acoustic Modeling was utilized to facilitate unsupervised learning of speech representations. Non-autoregressive predictive Coding (NPC)~\cite{liu2020non}, on the other hand, relies on the local dependencies of speech in a non-autoregressive manner to learn speech representations. This is accomplished by introducing Masked Convolution Blocks. HuBERT~\cite{hsu2021hubert} utilizes an offline clustering step to provide aligned target labels for a BERT-like prediction loss. A key element of this model is the application of the prediction loss only in the masked regions, compelling the model to learn the combination of acoustic and language models on continuous input. Wav2vec 2.0~\cite{baevski2020wav2vec} masks speech inputs in the latent space and addresses contrastive tasks defined by the quantization of jointly learned latent representations.

Although self-supervised models have achieved success, existing methods often overlook the incorporation of emotion-related information during the pre-training process. Emotions play a crucial role in human communication, and utilizing emotion-aware representations can significantly enhance the performance of emotion-related tasks, such as SER task. Specifically, emotional intensity knowledge in speech has been identified as a valuable factor that can further enrich the understanding of emotions \cite{liu2024emotion}. We believe that pre-training tasks incorporating emotion knowledge will contribute to a better understanding of emotions across the entire speech, thereby leading to improved performance in SER tasks.

In the field of NLP, certain studies strive to integrate both textual and emotional knowledge into pre-training models. For example, SentiLARE ~\cite{ke2019sentilare} injects word-level linguistic knowledge, such as part-of-speech tags and sentiment polarity, using the label-aware mask language model pre-training task to construct knowledge-aware language representations. Inspired by the renowned masked language modeling, eMLM~\cite{sosea2021emlm} introduces a novel emotion-related pre-training objective. Rather than masking tokens within the same input sequence indiscriminately, it leverages lexical information to assign higher masking probabilities to words that are more likely to be pivotal in emotional or affective contexts. In the field of speech, the Vesper~\cite{chen2024vesper} model introduces an enhanced emotion-specific pre-training encoder. It undergoes pre-training on a speech dataset based on WavLM~\cite{chen2022wavlm} while considering emotional features. To enhance sensitivity to emotional information, Vesper adopts an emotion-guided masking strategy to identify regions that require masking. However, the above works did not introduce intensity knowledge, which can aid pre-training models in acquiring more refined emotional information.

Introducing emotion intensity knowledge into a self-supervised pre-training model faces two difficulties: \begin{itemize}
    \item \textbf{Emotion intensity knowledge acquisition:} obtaining emotion intensity at the frame level from speech with emotions using models.
    \item \textbf{Pre-training task design:} incorporating the acquired emotion intensity information into the pre-training of self-supervised models.
\end{itemize}

To address the above challenges, we propose an emotion-aware speech representation learning method based on intensity knowledge. Our approach aims to elevate the performance of downstream tasks, particularly focusing on SER. Specifically, we first extract frame-level emotional intensity scores using established emotion extraction models. Secondly, by adopting a strategy inspired by the introduction of emotion in NLP text tasks, we propose an emotional masking strategy (EMS). This method instructs the self-supervised model to identify frames with heightened emotional intensity during the masking process. Applying our proposed method to Mockingjay~\cite{liu2020mockingjay} and NPC~\cite{liu2020non} models and utilizing the extracted advanced representations in SER tasks yields results that surpass the performance of the original models. The primary contributions of this work can be summarized as follows: 
\begin{itemize}
    \item We propose a novel emotion-aware speech SSL representation learning scheme.
    \item We extract fine-grained emotional intensity prior information from speech and propose emotional masking strategy. We select two classic SSL models, Mockingjay and NPC, as the subjects of our research.
    \item Experiments have demonstrated that the representations derived from our proposed method outperform the original model in SER task and other related tasks.
\end{itemize}

\begin{figure*}[ht]
  \centering
  \includegraphics[width=0.66\linewidth]{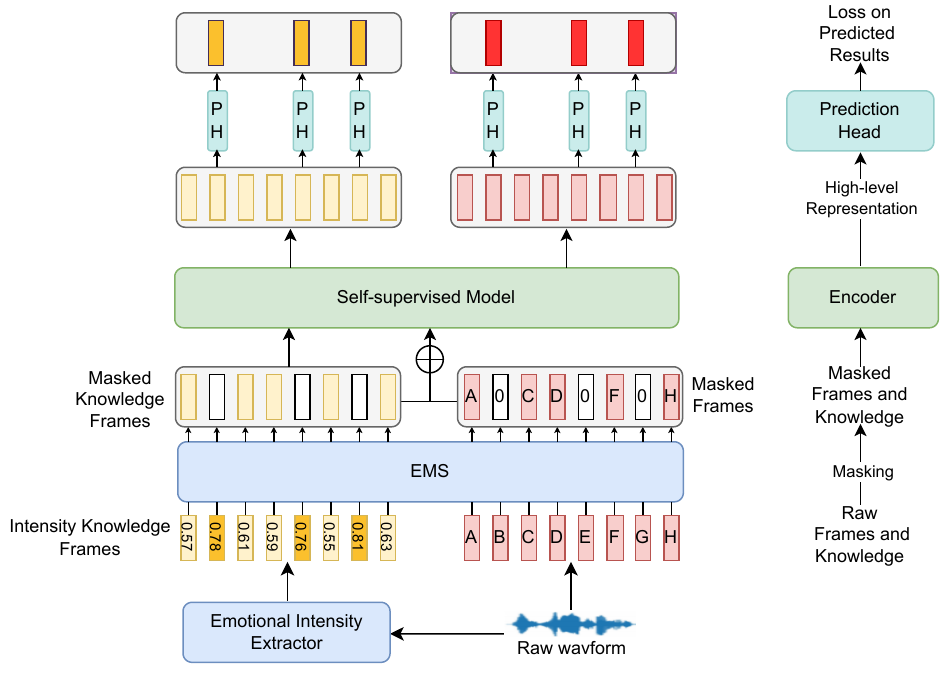} %
  \caption{Proposed model architecture. The small rectangles in the figure indicate the frame-level emotional intensity scores or acoustic features, those with numbers indicate emotional intensity scores, and the white parts indicate masked. The self-supervised model represents the encoder part of the improved model.}
  \label{fig:Model framework}
\end{figure*}

\section{Proposed Method}
In this section, we initially introduce the task definition and outline the overall workflow of the proposed method. Subsequently, we delve into the knowledge acquisition and training task sections. Finally, we elucidate how the EMS is applied to the Mockingjay~\cite{liu2020mockingjay} and NPC~\cite{liu2020non} models.

\subsection{Task Definition and Overall workflow}
Our task is defined as follows: given a speech sequence $X=\left[x_1, \cdots, x_T\right]$ of $T$ frames, our objective is to obtain high-level representations for the entire sequence, denoted as $H=\left(h_1, h_2, \cdots, h_n\right)^{\top} \in \mathbb{R}^{n \times d}$, where $d$ represents the dimension of the representation vectors. The goal is to capture both contextual and emotional knowledge.

As shown in Figure \ref{fig:Model framework}, our proposed method consists of two main components. Firstly, we employ an emotional intensity extractor to obtain the emotional intensity scores of the original audio. Subsequently, we use EMS to apply masks to both the acoustic frames and emotional intensity. Next, the masked results are fed into a self-supervised pre-training model. During training, the sum of the acoustic frames and emotional intensity scores is input to the model, and the resulting high-level representations are used for prediction through a prediction head. The L1 loss is applied to optimize the model, considering the prediction errors for both emotional intensity and the sum of emotional intensity and original frames. The overall loss is formulated as follows:
\begin{equation}
    \mathcal {L} = \mathcal{L}_{score} + \mathcal{L}_{joint\ input}
\end{equation}

\subsection{Knowledge Acquisition and Pre-training Task}
\subsubsection{Knowledge Acquisition}
This module is utilized to obtain the emotional intensity scores for each frame of the input audio. We employ Strengthnet~\cite{liu2021strengthnet} for this purpose, which comprises an acoustic encoder, an intensity predictor, and an emotion predictor.

The acoustic encoder consists of 12 convolutional layers that extract high-level features $H$ from the given input mel-spectrum sequence $X$. The high-level features $H$ are then fed into two predictors: one for predicting emotion intensity scores and the other for predicting emotion categories. The intensity predictor comprises a BiLSTM layer, two fully connected layers (FC), and an average pooling layer, which reads the high-level feature representation to predict emotion intensity scores. Similar to the intensity predictor, the emotion predictor includes a BiLSTM layer and an additional softmax layer. Using the encoder's output, the emotion predictor can forecast emotion categories.

As the predicted emotion intensity scores and the acoustic frames extracted by the feature extraction layer of the pre-trained self-supervised model may suffer from misalignment, we address this issue by applying a linear layer to align them before masking.

\subsubsection{Pre-training Task}
Given an input sequence of emotionally enhanced speech $X_k=\left\{\left(x_i,\text {score}_i\right)_{i=1}^n\right\}$, the objective of the pre-training task is to construct emotion-aware representation vectors $H=\left(h_1, h_2, \cdots, h_n\right)^{\top}$ that can facilitate downstream emotion-related tasks. We designed a novel supervised pre-training task called Emotional Masking  Strategy (EMS). This model incorporates frame-level emotional intensity scores during the pre-training phase to capture dependencies between utterance-level high-level representations and individual frames.

Unlike traditional masking strategies, such as the Masked Acoustic Modeling (MAM) proposed by Mockingjay~\cite{liu2020mockingjay}, which uses a uniform probability (15\%) to randomly mask acoustic frames, EMS employs a non-random masking approach. We assign higher probabilities to frames with higher emotion intensity scores. After calculating the intensity scores for each frame, we sorted them and selected a probability value of $k$ to mask the positions corresponding to the top $k$\% of the frames. The choice of value for $k$ is explained in detail in the experimental section.

\subsection{The pre-training model}
We selected Mockingjay~\cite{liu2020mockingjay} and NPC~\cite{liu2020non} as our pre-training models, whose core frameworks are Transformer and CNN, respectively. If our method showcases an enhancement in SER task performance across diverse frameworks, we contend that this underscores the effectiveness of emotion representations grounded in intensity knowledge.
\subsubsection{Mockingjay with EMS}
The Mockingjay~\cite{liu2020mockingjay} model utilizes a multi-layer Transformer encoder with multi-head self-attention~\cite{vaswani2017attention} for bidirectional encoding. Each encoder layer consists of two sub-layers: the first is a multi-head self-attention network, and the second is a feed-forward layer. Each sub-layer includes residual connections and layer normalization~\cite{ba2016layer}. The pre-training in Mockingjay is similar to the masked language modeling in BERT~\cite{devlin2018bert} and is conducted in a self-supervised setting. During the pre-training phase, continuous time steps from the encoder outputs are randomly masked. During training, the model adds a prediction head composed of a two-layer feed-forward network to predict the selected frames based on the left and right contextual information.

We made modifications to the masking strategy of Mockingjay~\cite{liu2020mockingjay}. During training, We select the top k\% of frames with the highest emotional intensity scores for masking. Similar to BERT \cite{devlin2018bert}, we introduce a sub-random process to improve training by addressing the mismatch between training and inference, where masked frames are absent during inference. This process involves three steps: 1) 80\% of the time, we mask the selected frames to zero, 2) 10\% of the time, we replace the selected frames with random frames, and 3) the remaining 10\% of the time, we leave the frames unchanged. Similar to the Mockingjay model, to prevent the model from exploiting the local smoothness of acoustic frames, we use additional consecutive masking. This forces the model to make inferences about global structure rather than relying on local information.

\subsubsection{NPC with EMS}

The NPC~\cite{liu2020non} model, to derive high-level features $h_t$ at time $t$ without global dependencies or autoregressive properties, restricts itself to depend only on the receptive field $\left(x_{t-r}, \ldots, x_t, \ldots, x_{t+r}\right)$ with a size of $R=2r+1$. In NPC, stacked convolutional blocks are used to build the representation extraction model. To ensure that the high-level feature $h_t$ indeed represents $x_t$, it is linearly transformed into $y_t$ to predict $x_t$. A vector-quantization~\cite{van2017neural} layer is employed as an information bottleneck before the linear projection to obtain better representations. The objective of NPC is to minimize the $L\text 1$ discrepancy between the surface feature $x_t$ and the prediction $y_t$ based on $h_t$ for all time steps
\begin{equation}
    \sum_{t=1}^T\left|y_t-x_t\right|
\end{equation}

To implement the desired restriction, NPC~\cite{liu2020non} introduces the Masked Convolution Block (Masked ConvBlock), where the kernel-wise convolution operation can be written as
\begin{equation}
    (W \odot D) * Z
\end{equation}
with $Z \in \mathbb{R}^{T \times d}$ denoting the intermediate features from model with sequence length $T$ and dimension $d, W \in \mathbb{R}^{k \times d}$ denoting the learnable kernel weight with size $k$, and $D \in\{0,1\}^{k \times d}$ denoting the mask with each element $d_{i j}=\mathbb{1}_{i \leq \frac{k}{2}-m}+\mathbb{1}_{i \geq \frac{k}{2}+m}$.

Similarly, we made modifications to the masking strategy of the NPC~\cite{liu2020non} model. After obtaining the emotion intensity scores for each frame, we determine which part of the convolutional kernel the higher emotion scores mainly appear in. Specifically, for each iteration of the acoustic frames with the size of the convolutional kernel, we record the emotion scores of each position traversed by the kernel. After completing the traversal, we identify the position in the kernel with the highest average emotion intensity score and set its corresponding weight to 0.

\section{Experiments and Results}
\subsection{Dataset}
\begin{table}[htbp]
    \centering
    \caption{Accuracy of the Mockingjay with EMS model on the SER task(\%).}
    
    \begin{tabular}{p{3.1cm}<{\centering}p{1.3cm}<{\centering}|p{1.5cm}<{\centering}}
        \toprule
        \multicolumn{2}{c|}{\textbf{Method}} & \textbf{ACC} $(\uparrow)$  \\
        \midrule
        \textbf{Mockingjay}&  & 50.28 \\
        \multirow{5}{*}{\textbf{Mockingjay with EMS}}& ($15\%$) & 55.94  \\
        & ($20\%$) & 55.76  \\
        & ($25\%$) & \textbf{57.42}  \\
        & ($30\%$) & 55.85  \\
        & ($35\%$) & 56.12  \\
        & ($40\%$) & 56.96  \\
        \bottomrule
    \end{tabular}
    \label{tab:table1}
\end{table}
We fine-tuned the modified models using the Interactive Emotional Dyadic Motion Capture (IEMOCAP)~\cite{busso2008iemocap} dataset consisting of approximately 12 hours of data, comprising 5 dyadic sessions performed by 10 professional actors. One session involves a conversation between two exclusive speakers.

During the SER evaluation, we followed the common evaluation protocol proposed by SUPERB~\cite{yang2021superb}, using the widely used SER dataset IEMOCAP. We excluded the unbalanced emotion categories and focused only on the neutral, happy, sad, and angry categories. The evaluation was conducted on the standard split five folds using cross-validation.

\subsection{Experimental Setup}
\textbf{Mockingjay} For the Mockingjay~\cite{liu2020mockingjay} model, we trained with the same parameters as in the original paper for a total of 950k steps. Subsequently, we fine-tuned the model for an additional 10k steps using the IEMOCAP~\cite{busso2008iemocap} dataset with different mask probabilities. The choice of mask probabilities ranged from the original 15\% and increased in 5\% increments up to 40\%.

\noindent\textbf{NPC} For the NPC~\cite{liu2020non} model, we trained with the same parameters as in the original paper for a total of 325k steps. Following that, we fine-tuned the model for an additional 10k steps using the IEMOCAP~\cite{busso2008iemocap} dataset. As NPC employs Masked Convolution Blocks for masking, with an original mask size of 5, we explored the impact of different mask probabilities on the experiments by selecting two additional mask sizes, 7 and 9, which were proposed to perform well in the original paper.

\subsection{Results on the SER Task}

Table \ref{tab:table1} compares the performance of Mockingjay~\cite{liu2020mockingjay} on the IEMOCAP~\cite{busso2008iemocap} dataset with different fine-tuning probabilities. The numbers in parentheses represent the masking probability. From the table, it can be observed that the results with masking probabilities ranging from 15\% to 40\% outperform Mockingjay's SER results. Specifically, the model fine-tuned with a 25\% masking probability demonstrates the best performance, achieving a 7.14\% improvement over the original model results.

Table \ref{tab:table2} compares the performance of NPC~\cite{liu2020non} on the IEMOCAP~\cite{busso2008iemocap} dataset with different mask sizes during fine-tuning. The numbers in parentheses represent the mask size. Additionally, we compared the impact of different inputs on the model during fine-tuning. "Separate Input" does not sum emotion intensity scores and acoustic frames; instead, they are separately input into the self-supervised model. "Joint Input" represents summing both and inputting the combined result and emotion intensity scores separately into the self-supervised model. From the table, it can be observed that both joint and separate inputs lead to improved accuracy for NPC in the SER task. Among them, joint input performs the best, achieving a 3.06\% accuracy improvement over the original model results. When the mask size is 7 and 9, the accuracy of the SER task decreases. We believe that a larger mask convolutional kernel may capture more emotional information but could lead to the loss of significant acoustic information.

\begin{table}[htbp]
    \centering
    \caption{Accuracy of the NPC model with EMS on the SER task(\%).}
    \begin{tabular}{p{3.5cm}<{\centering}|p{3.1cm}<{\centering}}
        \toprule
        \textbf{Method} & \textbf{ACC} $(\uparrow)$  \\
        \midrule
        \textbf{NPC} & 59.08 \\
        \textbf{NPC with EMS}($7$) & 47.10  \\
        \textbf{NPC with EMS}($9$) & 50.04  \\
        \textbf{Separate Input}($5$) & 60.56  \\
        \textbf{Joint Input}($5$) & \textbf{62.14}  \\
        \bottomrule
    \end{tabular}
    \label{tab:table2}
\end{table}

\begin{table}[htbp]
    \centering
    \caption{Results of the Mockingjay model with EMS on intention recognition and phoneme classification tasks(\%). IC denotes Intent Classification, PR denotes Phoneme Recognition, and FSC denotes Fluent Speech Commands dataset. LS denotes LibriSpeech dataset.}
    \begin{tabular}{p{3.9cm}<{\centering}|p{1.2cm}<{\centering}|p{1.2cm}<{\centering}}
        \toprule
        \textbf{Task} & \textbf{IC}  & \textbf{PR}   \\ \midrule
        \textbf{Dataset} & \textbf{FSC}  & \textbf{LS}   \\ \midrule
        \textbf{Model} & \textbf{ACC} $(\uparrow)$  & \textbf{PER} $(\downarrow)$ \\
        \midrule
        \textbf{Mockingjay} & 34.33 & 70.19 \\
        \textbf{Mockingjay with EMS}$(15\%)$ & 38.84 & \textbf{63.03}  \\
        \textbf{Mockingjay with EMS}$(25\%)$ & \textbf{45.35} & 63.38  \\
        \bottomrule
    \end{tabular}
    \label{tab:table3}
\end{table}

\subsection{Analysis on Generalization Ability}
\textbf{Generalization to Other Downstream Tasks:} In addition to the SER task, we also selected two other diverse tasks to explore whether our emotion-aware representation benefits them. These tasks include phoneme recognition, which focuses primarily on speech content, and intent classification, which emphasizes semantic content. The evaluation was conducted on the LibriSpeech~\cite{panayotov2015librispeech} train-clean-100/dev-clean/test-clean subsets and the Fluent Speech Commands~\cite{lugosch2019speech} dataset using the Mockingjay model trained with our approach.

The results in Table \ref{tab:table3} indicate that the extracted emotion-aware representation can enhance the performance of downstream tasks beyond those explicitly considering emotion-related objectives. In future work, we aim to explore the extension of our approach to a broader range of downstream tasks.

\section{Conclusion}

In this paper, we introduced an emotion-aware representation learning method based on intensity knowledge, called EMS. Extracting emotional intensity scores and designing a novel pre-training task injecting emotional knowledge into high-level representations, we enhanced the sentiment information in the representations. Through this approach, our method achieved significant performance improvements in SER tasks and demonstrated notable enhancements in other downstream tasks as well. In future work, we consider incorporating emotional knowledge at other granularities into high-level representations.

\section{Acknowledgement}

The research by Rui Liu was funded by the Young Scientists Fund of the National Natural Science Foundation of China (No. 62206136), Guangdong Provincial Key Laboratory of Human Digital Twin (No. 2022B121201 0004),
% the High-level Talents Introduction Project of Inner Mongolia University (No. 10000-22311201),
and the ``Inner Mongolia Science and Technology Achievement Transfer and Transformation Demonstration Zone, University Collaborative Innovation Base, and University Entrepreneurship Training Base'' Construction Project (Supercomputing Power Project) (No.21300-231510). 

\bibliographystyle{IEEEtran}
\bibliography{mybib}

% Generated by IEEEtran.bst, version: 1.13 (2008/09/30)
\begin{thebibliography}{10}
\providecommand{\url}[1]{#1}
\csname url@samestyle\endcsname
\providecommand{\newblock}{\relax}
\providecommand{\bibinfo}[2]{#2}
\providecommand{\BIBentrySTDinterwordspacing}{\spaceskip=0pt\relax}
\providecommand{\BIBentryALTinterwordstretchfactor}{4}
\providecommand{\BIBentryALTinterwordspacing}{\spaceskip=\fontdimen2\font plus
\BIBentryALTinterwordstretchfactor\fontdimen3\font minus
  \fontdimen4\font\relax}
\providecommand{\BIBforeignlanguage}[2]{{%
\expandafter\ifx\csname l@#1\endcsname\relax
\typeout{** WARNING: IEEEtran.bst: No hyphenation pattern has been}%
\typeout{** loaded for the language `#1'. Using the pattern for}%
\typeout{** the default language instead.}%
\else
\language=\csname l@#1\endcsname
\fi
#2}}
\providecommand{\BIBdecl}{\relax}
\BIBdecl

\bibitem{devlin2018bert}
J.~Devlin, M.-W. Chang, K.~Lee, and K.~Toutanova, ``Bert: Pre-training of deep
  bidirectional transformers for language understanding,'' \emph{arXiv preprint
  arXiv:1810.04805}, 2018.

\bibitem{radford2018improving}
A.~Radford, K.~Narasimhan, T.~Salimans, I.~Sutskever \emph{et~al.}, ``Improving
  language understanding by generative pre-training,'' 2018.

\bibitem{lan2019albert}
Z.~Lan, M.~Chen, S.~Goodman, K.~Gimpel, P.~Sharma, and R.~Soricut, ``Albert: A
  lite bert for self-supervised learning of language representations,''
  \emph{arXiv preprint arXiv:1909.11942}, 2019.

\bibitem{baevski2022data2vec}
A.~Baevski, W.-N. Hsu, Q.~Xu, A.~Babu, J.~Gu, and M.~Auli, ``Data2vec: A
  general framework for self-supervised learning in speech, vision and
  language,'' in \emph{International Conference on Machine Learning}.\hskip 1em
  plus 0.5em minus 0.4em\relax PMLR, 2022, pp. 1298--1312.

\bibitem{FAN2024102522}
\BIBentryALTinterwordspacing
C.~Fan, J.~Wang, W.~Huang, X.~Yang, G.~Pei, T.~Li, and Z.~Lv, ``Light-weight
  residual convolution-based capsule network for eeg emotion recognition,''
  \emph{Advanced Engineering Informatics}, vol.~61, p. 102522, 2024. [Online].
  Available:
  \url{https://www.sciencedirect.com/science/article/pii/S1474034624001708}
\BIBentrySTDinterwordspacing

\bibitem{liu2024contrastive}
R.~Liu, H.~Zuo, Z.~Lian, B.~W. Schuller, and H.~Li, ``Contrastive learning
  based modality-invariant feature acquisition for robust multimodal emotion
  recognition with missing modalities,'' \emph{IEEE Transactions on Affective
  Computing}, 2024.

\bibitem{fayek2017evaluating}
H.~M. Fayek, M.~Lech, and L.~Cavedon, ``Evaluating deep learning architectures
  for speech emotion recognition,'' \emph{Neural Networks}, vol.~92, pp.
  60--68, 2017.

\bibitem{vaswani2017attention}
A.~Vaswani, N.~Shazeer, N.~Parmar, J.~Uszkoreit, L.~Jones, A.~N. Gomez,
  {\L}.~Kaiser, and I.~Polosukhin, ``Attention is all you need,''
  \emph{Advances in neural information processing systems}, vol.~30, 2017.

\bibitem{zuo2023exploiting}
H.~Zuo, R.~Liu, J.~Zhao, G.~Gao, and H.~Li, ``Exploiting modality-invariant
  feature for robust multimodal emotion recognition with missing modalities,''
  in \emph{ICASSP 2023-2023 IEEE International Conference on Acoustics, Speech
  and Signal Processing (ICASSP)}.\hskip 1em plus 0.5em minus 0.4em\relax IEEE,
  2023, pp. 1--5.

\bibitem{busso2008iemocap}
C.~Busso, M.~Bulut, C.-C. Lee, A.~Kazemzadeh, E.~Mower, S.~Kim, J.~N. Chang,
  S.~Lee, and S.~S. Narayanan, ``Iemocap: Interactive emotional dyadic motion
  capture database,'' \emph{Language resources and evaluation}, vol.~42, pp.
  335--359, 2008.

\bibitem{livingstone2018ryerson}
S.~R. Livingstone and F.~A. Russo, ``The ryerson audio-visual database of
  emotional speech and song (ravdess): A dynamic, multimodal set of facial and
  vocal expressions in north american english,'' \emph{PloS one}, vol.~13,
  no.~5, p. e0196391, 2018.

\bibitem{liu2020mockingjay}
A.~T. Liu, S.-w. Yang, P.-H. Chi, P.-c. Hsu, and H.-y. Lee, ``Mockingjay:
  Unsupervised speech representation learning with deep bidirectional
  transformer encoders,'' in \emph{ICASSP 2020-2020 IEEE International
  Conference on Acoustics, Speech and Signal Processing (ICASSP)}.\hskip 1em
  plus 0.5em minus 0.4em\relax IEEE, 2020, pp. 6419--6423.

\bibitem{liu2020non}
A.~H. Liu, Y.-A. Chung, and J.~Glass, ``Non-autoregressive predictive coding
  for learning speech representations from local dependencies,'' \emph{arXiv
  preprint arXiv:2011.00406}, 2020.

\bibitem{hsu2021hubert}
W.-N. Hsu, B.~Bolte, Y.-H.~H. Tsai, K.~Lakhotia, R.~Salakhutdinov, and
  A.~Mohamed, ``Hubert: Self-supervised speech representation learning by
  masked prediction of hidden units,'' \emph{IEEE/ACM Transactions on Audio,
  Speech, and Language Processing}, vol.~29, pp. 3451--3460, 2021.

\bibitem{baevski2020wav2vec}
A.~Baevski, Y.~Zhou, A.~Mohamed, and M.~Auli, ``wav2vec 2.0: A framework for
  self-supervised learning of speech representations,'' \emph{Advances in
  neural information processing systems}, vol.~33, pp. 12\,449--12\,460, 2020.

\bibitem{liu2024emotion}
R.~Liu, Y.~Hu, Y.~Ren, X.~Yin, and H.~Li, ``Emotion rendering for
  conversational speech synthesis with heterogeneous graph-based context
  modeling,'' in \emph{Proceedings of the AAAI Conference on Artificial
  Intelligence}, vol.~38, no.~17, 2024, pp. 18\,698--18\,706.

\bibitem{ke2019sentilare}
P.~Ke, H.~Ji, S.~Liu, X.~Zhu, and M.~Huang, ``Sentilare: Sentiment-aware
  language representation learning with linguistic knowledge,'' \emph{arXiv
  preprint arXiv:1911.02493}, 2019.

\bibitem{sosea2021emlm}
T.~Sosea and C.~Caragea, ``emlm: a new pre-training objective for emotion
  related tasks,'' in \emph{Proceedings of the 59th Annual Meeting of the
  Association for Computational Linguistics and the 11th International Joint
  Conference on Natural Language Processing (Volume 2: Short Papers)}, 2021,
  pp. 286--293.

\bibitem{chen2024vesper}
W.~Chen, X.~Xing, P.~Chen, and X.~Xu, ``Vesper: A compact and effective
  pretrained model for speech emotion recognition,'' \emph{IEEE Transactions on
  Affective Computing}, 2024.

\bibitem{chen2022wavlm}
S.~Chen, C.~Wang, Z.~Chen, Y.~Wu, S.~Liu, Z.~Chen, J.~Li, N.~Kanda,
  T.~Yoshioka, X.~Xiao \emph{et~al.}, ``Wavlm: Large-scale self-supervised
  pre-training for full stack speech processing,'' \emph{IEEE Journal of
  Selected Topics in Signal Processing}, vol.~16, no.~6, pp. 1505--1518, 2022.

\bibitem{liu2021strengthnet}
R.~Liu, B.~Sisman, B.~Schuller, G.~Gao, and H.~Li, ``{Accurate Emotion Strength
  Assessment for Seen and Unseen Speech Based on Data-Driven Deep Learning},''
  in \emph{Proc. Interspeech 2022}, 2022, pp. 5493--5497.

\bibitem{ba2016layer}
J.~L. Ba, J.~R. Kiros, and G.~E. Hinton, ``Layer normalization,'' \emph{arXiv
  preprint arXiv:1607.06450}, 2016.

\bibitem{van2017neural}
A.~Van Den~Oord, O.~Vinyals \emph{et~al.}, ``Neural discrete representation
  learning,'' \emph{Advances in neural information processing systems},
  vol.~30, 2017.

\bibitem{yang2021superb}
S.-w. Yang, P.-H. Chi, Y.-S. Chuang, C.-I.~J. Lai, K.~Lakhotia, Y.~Y. Lin,
  A.~T. Liu, J.~Shi, X.~Chang, G.-T. Lin \emph{et~al.}, ``Superb: Speech
  processing universal performance benchmark,'' \emph{arXiv preprint
  arXiv:2105.01051}, 2021.

\bibitem{panayotov2015librispeech}
V.~Panayotov, G.~Chen, D.~Povey, and S.~Khudanpur, ``Librispeech: an asr corpus
  based on public domain audio books,'' in \emph{2015 IEEE international
  conference on acoustics, speech and signal processing (ICASSP)}.\hskip 1em
  plus 0.5em minus 0.4em\relax IEEE, 2015, pp. 5206--5210.

\bibitem{lugosch2019speech}
L.~Lugosch, M.~Ravanelli, P.~Ignoto, V.~S. Tomar, and Y.~Bengio, ``Speech model
  pre-training for end-to-end spoken language understanding,'' \emph{arXiv
  preprint arXiv:1904.03670}, 2019.

\end{thebibliography}

\end{document}